\documentclass[12pt]{article}

\newif\iffigs\figstrue

\usepackage{epsfig,latexsym}
\usepackage{amsmath}
\usepackage{color}
\usepackage{verbatim}
\usepackage{mathrsfs}
\usepackage{amssymb}

\DeclareMathAlphabet{\mathpzc}{OT1}{pzc}{m}{it}

 \csname
@addtoreset\endcsname{equation}{section}

\def\gz0{\gamma^{0}}

\def\scs#1{\section{\sc #1}}
\def\scss#1{\subsection{\sc #1}}

\def\a{\alpha}

\def\beq{\begin{equation}}
\newcommand{\eeq}[1]{\label{#1}\end{equation}}
\def\bea{\begin{eqnarray}}
\newcommand{\eea}[1]{\label{#1}\end{eqnarray}}
\def\ba{\begin{array}}
\def\ea{\end{array}}
\def\bec{\begin{center}}
\def\ec{\end{center}}
\def\ba{\begin{align}}
\def\ena{\end{align}}

\def\12{\frac{1}{2}}

\usepackage{slashed}

\usepackage{float}

\usepackage{booktabs}% for top-,mid- & bottomrule
\usepackage{caption}

\usepackage[mathscr]{euscript}

\usepackage{color}
\usepackage{graphicx}
\usepackage{epsf}
\usepackage{graphicx,epsfig}
\usepackage{amsmath}

\usepackage{multirow}

\usepackage{amsmath, amsthm, amssymb}

\usepackage{epsfig}
\usepackage{cite}
\usepackage{color,colordvi}

\newcounter{hran}

\makeatletter
\renewcommand\section{\@startsection {section}{1}{\z@}%
                               {-3.5ex \@plus -1ex \@minus -.2ex}%
                               {2.3ex \@plus.2ex}%
                               {\normalfont\large\bfseries}}
\makeatother

\vspace{0.5cm}

\newcommand{\bi}{\begin{itemize}}
\newcommand{\ei}{\end{itemize}}

\begin{document}
\thispagestyle{empty}
\begin{flushright}
CERN-PH-TH/2015-136\\
{\today}
\end{flushright}

\vspace{15pt}

\begin{center}

%%%%%%%%%%%%%%%%%%%%%%%%%%%%%%%%%%%%%%%%%%%%%%%%%%%%%%%%%%%%%%%%%%%%

{\Large\sc Some Pathways in non--Linear Supersymmetry}\\
{\sc Special Geometry Born--Infeld's, Cosmology and dualities}\\

%%%%%%%%%%%%%%%%%%%%%%%%%%%%%%%%%%%%%%%%%%%%%%%%%%%%%%%%%%%%%%%%%%%%

\vspace{35pt}
{\sc S.~Ferrara${}^{\; a,b,c*}$ and A.~Sagnotti${}^{\; a,d**}$}\\[15pt]

{${}^a$\sl\small Th-Ph Department, CERN\\
CH - 1211 Geneva 23, SWITZERLAND \\ }
\vspace{6pt}

{${}^b$\sl\small INFN - Laboratori Nazionali di Frascati \\
Via Enrico Fermi 40, I-00044 Frascati, ITALY}\vspace{6pt}

{${}^c$\sl\small Department of Physics and Astronomy, \\ U.C.L.A., Los Angeles CA 90095-1547,
USA}\vspace{6pt}

{${}^d$\sl\small Scuola Normale Superiore and INFN,\\
Piazza dei Cavalieri 7\
I-56126 Pisa, Italy}\vspace{6pt}

%%%%%%%%%%%%%%%%%%%%%%%%%%%%%%%%%%%%%%%%%%%%%%%
\vspace{8pt}

%%%%%%%%%%%%%%%%%%%%%%%%%%%%%%%%%%%%%%%%%%%%%%%
\vspace{25pt} {\sc\large Abstract}
\end{center}
\baselineskip=14pt
This review is devoted to some aspects of non--linear Supersymmetry in four dimensions that can be efficiently described via nilpotent superfields, in both rigid and curved Superspace. Our focus is mainly on the partial breaking of rigid $N=2$ Supersymmetry and on a class of generalized Born--Infeld systems that originate from Special Geometry and on some prototype cosmological models, starting from the Supergravity embedding of Starobinsky inflation. However, as an aside we also review briefly some interesting two--field extensions of the Born--Infeld Lagrangian whose field equations enjoy extended duality symmetries.

\vskip 12pt
\noindent DOI:
\vskip 24pt
{\sl
\noindent \small Contribution to the Proceedings of ``Group Theory, Probability, and the
Structure of Spacetime'', A Conference on the Occasion of Professor V.~S.~Varadarajan's Retirement, UCLA Mathematics Department, November 7--9, 2014. To appear in a  special issue of ``p-Adic Numbers, Ultrametric Analysis and Applications''.}
%%%%%%%%%%%%%%%%%%%%%%%%%%%%%%%%%%%%%%%%%%%%%%%
\vfill

\noindent
\baselineskip=20pt
\setcounter{page}{1}

\pagebreak

\newpage
%%%%%%%%%%%%%%%%%%%%%%%%%%%%%%%%%%%%%%%%%%%%%%%%%%%%%%%
\scs{Introduction}
%%%%%%%%%%%%%%%%%%%%%%%%%%%%%%%%%%%%%%%%%%%%%%%%%%%%%%%

This review is dedicated to prof. V.~S.~Varadarajan, with whom one of us (SF) has enjoyed pleasant and stimulating mutual interactions and collaborations over many years. It is meant to highlight some results related to non--linear realizations of Supersymmetry in both rigid and curved Superspace that were obtained by us and other authors during the last year. Our aim is to provide a readable account, focussing on a few conceptual lessons and leaving out a number of technical details that the interested reader can find in the original papers. This choice will allow us to touch upon several subjects whose leitmotif is indeed provided by non--linear realizations of Supersymmetry. Particular emphasis will be given to the role of nilpotent superfields \cite{nilpotent_1,nilpotent_2,nilpotent_3} in spontaneously broken Supersymmetry, for $N=1$ and also for $N=2$ partial breaking.

Non--linear realizations typically emerge when some members of a multiplet become particularly heavy and can be therefore integrated out at low energies. They typically bring along beautiful geometric structures \cite{ccwz}, a general fact whose key features were identified long ago and still play a pivotal role in our understanding of the Fundamental Interactions. What we shall add here to this well developed picture is a collection of examples where some generalizations of well--known structures emerge naturally from the partial breaking of rigid and local $N=2$ Supersymmetry \cite{partial_n2_0,partial_n2_1,partial_n2_2,partial_n2_3,partial_n2_4,partial_n2_5,
partial_n2_6,partial_n2_7,partial_n2_8}. On a different note, we shall also see how the coupling to Supergravity \cite{sugra} of these systems allows to formulate simple generalizations \cite{adfs} of the Starobinsky model \cite{starobinsky} which avoid some typical instabilities present in models with linearly realized Supersymmetry \cite{kallosh_linde}. While, as we have stressed, non--linear realizations have typically the flavor of low--energy approximations, they also appear to play a deeper role in String Theory \cite{stringtheory}. This occurs in models with ``Brane Supersymmetry Breaking'' (BSB) \cite{BSB}, orientifold vacua \cite{orientifolds} where a  high--scale breaking is induced by mutually non--supersymmetric collections of branes and orientifolds. While this setting brings non--linear realizations of Supersymmetry to the forefront at a potentially deeper level \cite{dm}, its four--dimensional counterparts in the low--energy Supergravity fall nicely into the class of models reviewed here \cite{fkl_1,fkl_2,fkl_3,dallz}, as we shall see in the following sections. Generalized continuous dualities of supersymmetric non--linear systems \cite{gz} of the Born--Infeld \cite{borninfeld,borninfeldsusy_1,borninfeldsusy_2,kt} type are a final, related aspect that we shall touch upon. As is well known, the equations of motion of the Born--Infeld theory are invariant under a continuous $U(1)$ duality, in analogy with their simpler counterparts in the Maxwell theory. On the other hand, an interesting class of more general Abelian models \cite{fps,adt_15} that we shall deduce from $N=2$ rigid Special Geometry \cite{specialgeometry_1,specialgeometry_2} does not enjoy, in general, a similar property. We shall therefore conclude with a simple illustration of three types of two--field extensions of the Born--Infeld theory whose field equations are invariant under $U(1)$, a $U(1) \times U(1)$ and $SU(2)$ continuous duality rotations \cite{kt,abmz,fsy}.

The review opens in Section \ref{sec:2} with a brief introduction to (partial) Supersymmetry breaking in $N$--extended rigid Supersymmetry, focussed on the emergence of non--linear actions in certain limits. Following \cite{nilpotent_1,nilpotent_2,nilpotent_3}, these are best captured in terms of nilpotent superfields where some bosonic degrees of freedom are eliminated, which are reviewed in Section \ref{sec:3} and play a central role in the prototype Volkov--Akulov \cite{va} and Born--Infeld systems \cite{borninfeld,borninfeldsusy_1,borninfeldsusy_2,kt}. In Section \ref{sec:4} we review multi--field generalizations of the Born--Infeld theory that obtain as effective low--energy descriptions of $N=2$ systems with Fayet--Iliopoulos terms \cite{FI,partial_n2_5}, in the limit where $N=1$ chiral sub--multiplets of complete $N=2$ vector multiplets become very massive and can be integrated out \cite{partial_n2_8,fps}. In section \ref{sec:5} we briefly elaborate on analogies between these systems and $N=2$ extremal black holes \cite{attractors}, and then in Section \ref{sec:6} we discuss the solutions of the non--linear constraints derived in Section \ref{sec:4} that identify explicit multi--field generalizations of the Born--Infeld system for the cases of two and three vector multiplets \cite{fps}, making use of some results in the theory of invariant polynomials.
Section \ref{sec:7} is devoted to some realizations of non--linear supersymmetry in Supergravity, and in particular to the Starobinsky model of inflation \cite{adfs} and to a few other significant inflaton models based on nilpotent superfields \cite{fkl_1,fkl_2,fkl_3,dallz}. In Section \ref{sec:8} we review the construction of non--linear systems for massless $p$--forms and their massive variants \cite{fs15,fsy}. We conclude in Section \ref{sec:9} with some comments on the continuous dualities of the field equations for these non--linear Lagrangians, which extend the familiar property of the Maxwell theory. While the multi--field generalizations of the Born--Infeld system discussed in Sections \ref{sec:4} and \ref{sec:6} typically do not possess these continuous symmetries, we finally exhibit, in Section \ref{sec:9}, three interesting two--field extensions that do \cite{kt,abmz,fsy}, as an instructive application of the Gaillard--Zumino formalism \cite{gz}.

%%%%%%%%%%%%%%%%%%%%%%%%%%%%%%%%%%%%%%%%%%%%%%%%%%%%%%%
\scs{Spontaneous supersymmetry breaking}\label{sec:2}
%%%%%%%%%%%%%%%%%%%%%%%%%%%%%%%%%%%%%%%%%%%%%%%%%%%%%%%

In this section we give some generalities on the spontaneous breaking of rigid Supersymmetry \cite{partial_n2_0,partial_n2_1,partial_n2_2,partial_n2_3,partial_n2_4,partial_n2_5,
partial_n2_6,partial_n2_7,partial_n2_8}. To begin with, as is the case for any action invariant under continuous symmetries, in theories invariant under one or more Supersymmetry transformations there are conserved Majorana (vector--spinor) Noether currents \cite{fz} (here in van der Waerden notation):
\beq
\partial^\mu \, J_{\mu\, \alpha}^A(x) \ = \ 0 \quad (A=1,..,N) \ .
\eeq{susyb1}
The corresponding charges
\beq
Q_\alpha^A \ = \ \int d^3 {\bf x} \, J_{0\, \alpha}^A(x)
\eeq{susyb2}
generate the (N--extended) supersymmetry algebra
\beq
\left\{Q_\alpha^A\, , \, \overline{Q}_{\dot{\alpha} \ B} \right\}\ = \ \left(\sigma_\mu\right)_{\alpha\dot{\alpha}}\, P^\mu \ \delta^A_B \ , \quad \left\{Q_\alpha^A\, , \, Q_\beta^B \right\} \ = \ 0 \ .
\eeq{susyb3}
$p$--branes can modify this algebra by the addition of central extension terms \cite{town,fp}:
\begin{eqnarray}
&& \left\{Q_\alpha^A\, , \, \overline{Q}_{\dot{\alpha} \ B} \right\}\ = \ \left(\sigma_\mu\right)_{\alpha\dot{\alpha}}\, P^\mu \ \delta^A_B \ + \ Z^{A\, \mu}_B \ \left( \sigma_\mu\right)_{\alpha \dot{\alpha}}  \ ,\label{susyb41} \\
&& \left\{Q_\alpha^A\, , \, Q_\beta^B \right\} \ = \ Z^{[AB]}\, \epsilon_{\alpha\beta} \ + \ Z^{(AB)}_{(\alpha\beta)} \ . \label{susyb4}
\end{eqnarray}
In particular, the central term in eq.~\eqref{susyb41} corresponds to strings and the one in eq.~\eqref{susyb4} corresponds to points and membranes.

In the case of $N=1$ spontaneous breaking, the order parameter enters a term linear in the supercurrent \cite{salam,FI},
\beq
J_{\mu \, \dot{\alpha}} \ = \ f \, \left(\gamma_\mu G\right)_{\dot{\alpha}}
 + \ ... \ ,
\eeq{susyb5}
where $G$ is the Goldstino field and $f$ has dimension two in natural units.
As is generally the case for Goldstone particles, $G$ is a massless fermion whose low--energy self--interactions are described by a Lagrangian invariant under a non--linear realization of Supersymmetry.

When the central extensions vanish, the spontaneous breaking can only occur collectively \cite{partial_n2_0}, for all spinor charges at once and with the same breaking order parameter $f$. The low--energy effective Lagrangian for $N$ Goldstino fields is an obvious extension of the Volkov--Akulov action \cite{va} derived in the $N=1$ case, and in general reads
\beq
{\cal L}_{VA}\left(f,G_\alpha^A \right) \ = \ f^2 \left[ 1 \ - \ \sqrt{\,-\,{\rm det}\left(\eta_{\mu\nu} \ + \ \frac{i}{f^2}\ \left( \overline{G}_A \gamma_\mu \partial_\nu \,G^A - h.c.\right)\right)}\right] \ .
\eeq{susyb6}
This Goldstino Lagrangian contains a finite number of terms due to the nilpotency of $G$. It possesses a $U(n)$ R--symmetry, and is invariant under the non--linear supersymmetry transformations
\beq
\delta\, G_\alpha^A(x) \ = \ f\, \epsilon_\alpha^A \ + \ \frac{i}{f}\, \left(\overline{G}_B \,\gamma^\mu \epsilon^B \ - \ \overline{\epsilon}_B \,\gamma^\mu G^B \right) \ \partial_\mu\, G_\alpha^A \ .
\eeq{susyb7}

The Lagrangian \eqref{susyb6} is a consequence of the Supersymmetry current algebra, which in the absence of central extensions reads \cite{fz}
\beq
\langle 0| \left\{ \overline{Q}_{\dot{\alpha}\,A},J_{\mu\alpha}^B(x) \right\} |\rangle \ = \ \left( \sigma^\nu\right)_{\alpha \, \dot{\alpha}}\ \langle 0| T_{\nu\mu}| 0 \rangle \, \delta_A^B \ = \ \left( \sigma_\mu\right)_{\alpha \, \dot{\alpha}}\  f^2 \, \delta_A^B \ ,
\eeq{susyb8}
and the corresponding linear term in the supercurrent is
\beq
J_{\mu \, \dot{\alpha}\,A}\ = \ f \, \left(\gamma_\mu G_A\right)_{\dot{\alpha}}
 + \ ...
\eeq{susyb9}
One can invalidate this result modifying the current algebra by terms not proportional to $\delta_A^B$, adding a contribution proportional to a constant matrix $P_A^B$ valued in the adjoint of $SU(N)$, \emph{i.e} traceless and Hermitian \cite{partial_n2_5,partial_n2_4}:
\beq
\langle 0| \left\{ \overline{Q}_{\dot{\alpha}\,A},J_{\mu\alpha}^B(x) \right\} |0\rangle \ = \ \left( \sigma^\nu\right)_{\alpha \, \dot{\alpha}}\ \langle 0|  T_{\nu\mu} | 0\rangle \, \delta_A^B \ + \ \left( {\sigma_\mu} \right)_{\alpha\,\dot{\alpha}}\, P_A^B \ .
\eeq{susyb10}
Now the effective Goldstino actions depend on both $N$ and $P_A^B$, since it can be shown that
\beq
\delta_A \chi^i \, \delta^B {\overline{\chi}}_i \ = \ V\, \delta_A^B \ + \ P_A^B \quad \left( \delta \chi^i \ = \ \delta_A \chi^i \, \epsilon^A\right) \ .
\eeq{susyb11}
If $k$ Supersymmetries are unbroken, the Hermitian matrix $V \bf{1} + P$ has rank $N-k$ in the vacuum. Therefore, the $N-1$ possible scales of partial Supersymmetry breaking are expected to the classified by the $N-1$ Casimir operators of $SU(N)$. As is well known, algebraic equations can be generally solved in terms of radicals up to the quartic order, which amusingly corresponds in this setting to $N=4$, the maximal value of $N$ for rigid Supersymmetry.

A rich class of field theoretical examples is available for $N=2$, and the previous analysis in terms of $SU(2)$ invariants was performed in \cite{adft_15}. In this case the $SU(2)$ quadratic invariant is the squared norm of a 3--vector $\xi^x$ constructed via electric and magnetic Fayet--Iliopoulos (FI) terms $Q_x = (m_x^\Lambda,e_{x\,\Lambda})$,
\beq
\xi^x = \left( {\cal Q}_y \wedge {\cal Q}_z \right)\epsilon^{xyz} \ , \quad  {\cal Q}_y \wedge {\cal Q}_z  = m_y^\Lambda \, e_{z\Lambda} \ - \ m_z^\Lambda \, e_{y\Lambda} \ ,
\eeq{susyb12}
where $\Lambda = 1 , \ldots , n$ and $n$ is the number of vector multiplets. The wedge product is with respect to the symplectic structure inherited from the rigid Special Geometry, and the $P_A^B$ matrix takes the explicit form
\beq
\delta_A \chi^i \, \delta^B \overline{\chi}_i \ = \ V\, \delta_A^B \ + \ \left(\sigma_x\right)_A^B \, \xi^x \ = \ \left(\begin{array}{cc} V-\xi^3 & \xi^1+i\, \xi^2 \\ \xi^1-i\, \xi^2 & V + \xi^3\end{array} \right)  \ .
\eeq{susyb125}

We anticipate that the Goldstino action for partially broken $N=2$ Supersymmetry is described by the supersymmetric Born--Infeld theory, which rests on a non--linear Lagrangian ${\cal L}_{SBI}\left(G_\alpha,F_{\mu\nu} \right)$ for an $N=1$ vector multiplet whose chiral superfield strength $W_\alpha=\overline{D}^{\,2} \,D_\alpha\,V$ contains the goldstino $G$ at $\theta=0$ and the self--dual Maxwell field strength at the next order in $\theta$, with the two properties
\begin{eqnarray}
&& {\cal L}_{SBI}\left(G_\alpha, F_{\mu\nu}=0 \right) \ \longrightarrow \ {\cal L}_{VA} \ . \nonumber \\
&& {\cal L}_{SBI}\left(G_\alpha=0, F_{\mu\nu} \right) \ = \ f^2 \left[1 \ - \ \sqrt{\ -\ det\left( \eta_{\mu\nu} \ + \ \frac{1}{f} \ F_{\mu\nu}\right)} \right] \ . \label{susyb13}
\end{eqnarray}

This non--linear Lagrangian for the partial breaking $N=2 \to N=1$ enjoys two types of super--invariances: a manifest $N=1$ Supersymmetry, which is linearly realized in $N=1$ Superspace, and a second Supersymmetry which is non--manifest and non--linearly realized. In terms of microscopic parameters, the scale of the broken Supersymmetry is an $SU(2)$ and symplectic invariant quantity,
\beq
f^\frac{1}{2} \ = \ \left( \xi^x \, \xi^x \right)^\frac{1}{4} \ .
\eeq{susyb14}
The Goldstino action for a partial breaking $N \to N-k$ is only known for $N=2$, which corresponds to the non--linear limit of a quadratic action of vector multiplets endowed with
Fayet--Iliopoulos terms. In the simplest case of an $N=2$ vector multiplet, composed of an $N=1$ vector multiplet $W_\alpha$ and an $N=1$ chiral multiplet $X$, according to
\beq
\left(1,2\left(\frac{1}{2}\right),0,0\right) \ \longrightarrow \ \left(1,\frac{1}{2}\right) \ + \ \left(\frac{1}{2},0,0\right) \ ,
\eeq{susyb15}
the $P_A^B$ allow to give a mass $m_X$ to the chiral multiplet $X$, and in the large mass limit $X$ can be integrated out, giving rise to a non--linear theory for the fields $G_\alpha$ and $F_{\mu\nu}$ of the $N=1$ vector multiplet \cite{fps}.

%%%%%%%%%%%%%%%%%%%%%%%%%%%%%%%%%%%%%%%%%%%%%%%%%%%%%%%
\scs{Nilpotency constraints in $N=1$ Superspace}\label{sec:3}
%%%%%%%%%%%%%%%%%%%%%%%%%%%%%%%%%%%%%%%%%%%%%%%%%%%%%%%

Both the Goldstino Volkov--Akulov Lagrangian and the supersymmetric Born--Infeld Lagrangian follow from algebraic constraints in Superspace among various multiplets \cite{nilpotent_2,nilpotent_3}. The $N=1$ Volkov--Akulov Lagrangian is thus given by
\beq
{\cal L} \ = \ \left. X \, \overline{X}\right|_D \ +  \ \left. f\, X \right|_F \ ,
\eeq{susyb16}
where $X$ is a chiral superfield ($\overline{D}_{\dot{\alpha}} X =0$), subject to the nilpotency constraint $X^2=0$, whose solution is
\beq
X \ = \ \frac{G\,G}{F_G} \  + \ i \, \sqrt{2}\, \theta\, G \ + \ \theta^2 \, F_G \ .
\eeq{susyb17}
Eliminating the auxiliary field $F_G$, this theory can be turned into the Volkov--Akulov Lagrangian of eq.~\eqref{susyb6}.

In a similar fashion, the superfield constraint giving rise to the supersymmetric Born--Infeld action enforces a non--linear relation between the two $N=1$ supermultiplets $X$ and $W_\alpha$ that build the $N=2$ vector multiplet:
\beq
{W^2} \ + \ X\left( m_1\ - \ \bar{D}^2 \bar{X}\right) \ = \ 0 \ ,
\eeq{susyb18}
This was first identified in \cite{partial_n2_6}, and $m_1$ corresponds to a particular choice of a magnetic FI term \cite{partial_n2_8}. The non--linear relation \eqref{susyb18} determines $X$ as a non--linear function of $D^2\,W^2$ and $\overline{D}^{\,2}\,\overline{W}^{\,2}$ and implies the nilpotency constraints
\beq
X^2 \ = \ 0 \ , \quad  X \, W_\alpha \ = \ 0 \ .
\eeq{susyb1856}
The Born--Infeld action then takes the form
\beq
{\cal L}_{SBI} \ = \ Im \, \int d^2 \theta \ (e_1+ie_2) \ X\left(m_1, D^2\,W^2,\overline{D}^{\,2}\,\overline{W}^{\,2}\right) \ ,
\eeq{susyb19}
where $e_1$ and $e_2$ are electric FI terms whose labels correspond to the first two directions in an $SU(2)$ triplet. For canonically normalized vectors, one can then recognize that the scale $f$ and the theta--angle $\vartheta$ can be expressed in terms of $m_1$, $e_1$ and $e_2$ as
\beq
f \ = \ \sqrt{m_1\,e_2} \ , \quad \vartheta \ = \ \frac{e_1}{e_2} \ .
\eeq{susyb20}
Note that we chose a vanishing FI term along the third direction, since this guarantees that the matrix $P_A^B$ is diagonal, so that the unbroken Supersymmetry can be identified with the one manifest in Superspace.

%%%%%%%%%%%%%%%%%%%%%%%%%%%%%%%%%%%%%%%%%%%%%%%%%%%%%%%
\scs{$N=2$ rigid Special Geometry and FI terms}\label{sec:4}
%%%%%%%%%%%%%%%%%%%%%%%%%%%%%%%%%%%%%%%%%%%%%%%%%%%%%%%

We can now generalize the standard supersymmetric Born--Infeld model to the multi-field case, in a way dictated by Special Geometry and the partial $N=2 \to N=1$ breaking of Supersymmetry. To this end, let us first identify the two key data of the problem. To begin with, there are $N=2$ FI terms that build up an $Sp(2n)$ symplectic triplet of electric and magnetic charges $Q_x\,=\,({m_x}^\Lambda,\,e_{x \Lambda})$, with $x=1,2,3$,  $\Lambda=1,..,n$. Moreover, there is a holomorphic prepotential,
\beq
U(X)\ = \ \frac{i}{2} \ C_{\Lambda\Sigma} \, X^\Lambda \, X^\Sigma \ + \ \frac{1}{3!\,M} \ d_{\Lambda\Sigma\Gamma}\, X^\Lambda \,X^\Sigma \, X^\Gamma \ ,
\eeq{13}
where $C_{\Lambda\Sigma}$ and $d_{\Lambda\Sigma\Gamma}$ are totally symmetric and real. Here we have exhibited the dominant terms for large values of $M$, a mass parameter that sets the scale of the problem, but for brevity in the following we shall set $M=1$.

Eq.~(\ref{13}) clearly identifies the $d_{ABC}$ as third derivatives of the prepotential $U$. Moreover, the $N=2$ Lagrangian with an $N=2$ FI term, written in $N=1$ language, acquires a symplectic structure due to the underlying Special Geometry, which is encoded in the symplectic vector~\cite{specialgeometry_1,specialgeometry_2}
\beq
{\cal V} \ = \ \left(X^\Lambda, \ U_\Lambda\, \equiv \, \frac{\partial U}{\partial X^\Lambda}\right) \ .
\eeq{15}
The construction of the terms quadratic in the vector fields rests of on a pair of $n\times n$ symmetric matrices $g_{\Lambda\Sigma}$ and $\theta_{\Lambda\Sigma}$, which depend in general on the scalar fields,
\beq
{\cal L}\ = \ - \ \frac{1}{4} \ g_{\Lambda\Sigma}\, F_{\mu\nu}^\Lambda \, F^{\Sigma\, \mu\nu} \ +\  \frac{1}{8} \ \theta_{\Lambda\Sigma} \, F^\Lambda_{\mu\nu} \, F^\Sigma_{\rho\sigma} \ \epsilon^{\mu\nu\rho\sigma}\ ,
\eeq{17}
where the $F^\Lambda_{\mu\nu}$ are the gauge field strengths.

In $N=2$ Special Geometry
\beq
g_{\Lambda\Sigma}\ =\ Im \,U_{\Lambda\Sigma}\ , \qquad \theta_{\Lambda\Sigma} \ =\ Re \, U_{\Lambda\Sigma}\ , \qquad U_{\Lambda\Sigma}\ = \ \frac{\partial^{\,2} \, U}{\partial X^\Lambda \partial X^\Sigma} \ ,
\eeq{17a}
and it is convenient to define the symplectic metric
\beq
 \Omega \ = \ \left( \begin{matrix} 0 & -1 \\ 1 & 0 \end{matrix} \right) \ .
\eeq{17b}
The $2n\times 2n$ matrix ${\cal M}$, with entries
\beq
{\cal M} \ = \ \left( \begin{matrix} g \ + \ \theta \,g^{\,-1} \, \theta & -\,\theta \, g^{\,-1} \\ -\, g^{\,-1} \, \theta & g^{\,-1} \end{matrix} \right)\ ,
\eeq{17c}
then satisfies the two conditions of being symplectic and positive definite:
\beq
\qquad
{\cal M}\  =\  {\cal M}^{\,T} \ , \qquad {\cal M}\, \Omega \, {\cal M}\ = \ \Omega \ ,
\eeq{17d}
for a positive definite $g$, as required by the Lagrangian terms in eq.~\eqref{17}.

The contributions to the potential involve the symplectic triplets $Q_x=(m_x^\Lambda,\,e_{x\,\Lambda})$ of electric and magnetic charges. The first two combine naturally into the complex sets
\beq
Q \ \equiv \ (m^\Lambda \,, \, e_\Lambda) \ =\ (m_1^\Lambda \ + \ i\, m_2^\Lambda\,, \, e_{\,1 \Lambda} \ + \ i\, e_{\,2 \Lambda})
\eeq{19}
and determine the superpotential
\beq
{\cal W} \ = \  {\cal V}^{\,T} \, \Omega \ Q \ = \ \left( U_\Lambda \ m^\Lambda \ - \ X^\Lambda \ e_\Lambda \right) \ .
\eeq{17e}
The last,
\beq
Q_3\ = \left(m_3^\Lambda\, ,\ e_{3 \Lambda}\right)\ ,
\eeq{19a}
is real and determines, in $N=1$ language, the magnetic and electric FI $D$--terms. The potential of the theory can thus be expressed, in $N=1$ language, as
\beq
V \ = \  V_F \ + \ V_D  \ , \
\eeq{19aa}
where
\bea
V_F &=& (Im \, U^{-1} )^{\Lambda\Sigma} \ \frac{\partial \,{\cal W}}{\partial X^\Lambda}\ \frac{\partial \, {\cal W} }{\partial X^\Sigma} \  = \
\bar{Q}^T ({\cal M} \ - \ i\, \Omega ) Q \ , \label{21} \\
V_D &=& Q^T_3\, {\cal M} \, Q_3 \ .
\eea{22}

Vacua preserving an $N=1$ supersymmetry aligned with the $N=1$ superspace \cite{partial_n2_8} are determined by critical points of the potential \footnote{Notice that this is not the case in the original model of \cite{partial_n2_5} where the $D$-term
has  a non--vanishing VEV, so that the unbroken supersymmetry is a mixture of the two original $N=2$ superspace supersymmetries. The $N=2$ $SU(2)$ R--symmetry allows in fact to rotate $N=1$ FI terms into superpotential terms.}, and thus by the attractor equations
\beq
\frac{\partial \,{\cal W}}{\partial X^\Lambda}\ =\ 0 \ ,
\eeq{23}
which are in this case \cite{fps}
\beq
\left({\cal M} \ - \ i\, \Omega\right)Q\ =\ 0\ .
\eeq{24}
More explicitly, eqs.~\eqref{23} and \eqref{24} result in a condition for the vacuum expectation values $x^\Lambda = \langle X^\Lambda \rangle$, which takes the form
\beq
U_{\Lambda\Sigma}(x)\ m^\Sigma\  \equiv \ \left(i\,C_{\Lambda\Sigma}\ + \ d_{\Lambda\Sigma\Gamma}\ x^\Gamma\right)m^\Sigma\ = \ e_\Lambda\ ,
\eeq{5a}
with $m^\Sigma$ real, $e_\Lambda\ =\ e_{1 \Lambda} \ + \ i\,e_{2 \Lambda}$ and $e_{3 \,\Lambda} = m^{3 \,\Lambda} = 0$ in order to exclude $D$--term contributions, which would misalign the unbroken supersymmetry with respect to the $N=1$ superspace, as in \cite{partial_n2_5}. Eq.~\eqref{5a} implies the two real equations
\beq
\left(C_{\Lambda\Sigma}\ + \ d_{\Lambda\Sigma\Gamma}\, Im \, x^\Gamma\right)m^\Sigma\ =\ e_{2 \,\Lambda} \ , \qquad d_{\Lambda\Sigma\Gamma}\ Re \,x^\Gamma\, m^\Sigma \ =\ e_{1 \,\Lambda} \ .
\eeq{6a}
A non--vanishing $C_{\Lambda\Sigma}$ is needed to restore positivity of the kinetic term whenever the matrix $d_{\Lambda\Sigma\Gamma}\, m^\Gamma$ is not positive definite.

These equations can admit a solution for nonzero $Q$ only if
\beq
i\, \bar{Q}^T\, \Omega \, Q \ = \ i\left( m^\Lambda \, \bar{e}_\Lambda \ - \ \bar{m}^\Lambda\, e_\Lambda \right) \ > \ 0 \ ,
\eeq{24a}
since at the critical point ${\cal M}$ is positive definite, while the condition $V_D=0$ implies $Q_3=0$. All in all, with our choices eq.~\eqref{24a} reduces to the condition
\beq
m^\Lambda\, {e}_{\,2\,\Lambda} \ > \ 0 \ .
\eeq{24b}

The complete Lagrangian reads
\beq
{\cal L} \ = \ - \ Im \, \int d^2 \theta \, \left[U_{\Lambda\Sigma}\,W^\Lambda\,W^\Sigma \ + \ {\cal W}(X) \ + \ \frac{1}{2} \  \bar{D}^2 \, \left(X^\Lambda\, \bar{U}_\Lambda\ -\ \bar{X}^\Lambda\,  U_\Lambda\right)\right] \ ,
\eeq{29}
and defining the shifted fields $Y^\Lambda$ via $X^\Lambda = x^\Lambda + Y^\Lambda$, the Euler--Lagrange equations for the $Y^\Lambda$ take the form
\beq
d_{\Lambda\Sigma\Gamma}\left[ W^\Sigma\, W^\Gamma + Y^\Sigma\left( m^\Gamma - \bar{D}^2 \bar{Y}^\Gamma\right) \, + \, \frac{1}{2} \ \bar{D}^2 \left( \bar{Y}^\Sigma\, \bar{Y}^\Gamma\right)\right] \, + \, \left[\bar{U}_{\Lambda\Sigma}(x) - U_{\Lambda\Sigma}(x)\right]\bar{D}^2 \bar{Y}^\Sigma \, = \, 0 \ .
\eeq{30a}
Notice that the last two contributions are total derivatives, and therefore can be neglected in the infrared limit, where these equations reduce to
\beq
d_{\Lambda\Sigma\Gamma}\left[ W^\Sigma\, W^\Gamma + Y^\Sigma\left( m^\Gamma - \bar{D}^2 \bar{Y}^\Gamma\right) \right] \ = \ 0 \ ,
\eeq{30ab}
together with the nilpotency constraints
\beq
d_{\Lambda\Sigma\Gamma}\, Y^\Sigma \, Y^\Gamma \ = \ 0 \ , \quad d_{\Lambda\Sigma\Gamma}\, Y^\Sigma \, W^\Gamma \ = \ 0 \ .
\eeq{30abc}
The three equations \eqref{30ab} and \eqref{30abc}, which are manifestly $N=1$ supersymmetric, are in fact rotated into one another by the second non--linearly realized supersymmetry \cite{partial_n2_6}
\bea
\delta \, W^\Lambda_\a &=& m^\Lambda\, \eta_\a \ - \ \bar{D}^{\,2} \, \bar{X}^\Lambda \, \eta_\a \ - \ 4\, i\,  \partial_{\a\bar{\a}} \, X^\Lambda \ {\bar{\eta}}^{\,\bar{\a}} \ , \label{3a} \\
\delta \, X^\Lambda &=& - \ 2\ W^{\Lambda \, \a} \ \eta_\a \ .
\eea{4aaa}

The $\theta^2$ component of eq.~\eqref{30ab},
\beq
d_{\Lambda\Sigma\Gamma} \left[ F_{+}^\Sigma \cdot F_{+}^\Gamma \ + \  {\cal F}^\Sigma \left( m^\Gamma \, -\, \bar{\cal F}^\Gamma \right)  \right] \ = \ 0 \ ,
\eeq{30d}
where $F_{+}^\Gamma$ are self--dual field strength combinations and ${\cal F}^\Gamma$ is the $Y^\Gamma$ auxiliary field,  is the multi--field generalization of the Born--Infeld constraint \eqref{susyb18} that is induced by the Special Geometry.

The non--linear superspace Lagrangian is in general
\beq
{\cal L} \ = \ Im \int d^2 \theta \ e_\Lambda \, Y^\Lambda \ - \ Re \, \int d^2 \theta \ C_{\Lambda\Sigma} \left[ W^\Lambda\, W^\Sigma \ + \ Y^\Lambda \left( m_1^\Sigma \ - \ \bar{D}^2\, \bar{Y}^\Sigma \right) \right] \ ,
\eeq{12aa}
and $n=1$ its bosonic part reduces to
\begin{eqnarray}
{\cal L} &=& Im \int d^2 \theta \ e \, Y \ = \  \bigg( e_1 \, Im \, {\cal F} \ + \ e_2 \, Re \, {\cal F} \bigg) \nonumber \\ &=&
- \, \vartheta \ F \cdot \widetilde{F} \ + \ \frac{f^2}{2}\, \left[ 1 \, - \, \sqrt{1 \, + \,
\frac{4}{f^2} \ F \cdot F \, - \, \frac{4}{f^4} \ \left(  F \cdot \widetilde{F} \right)^2}\ \right]\, , \label{30e}
\end{eqnarray}
in terms of a canonically normalized field strength, obtained via the redefinition $F \to \sqrt{\frac{m}{e_2}} \, F$, where $\vartheta$ and $f$ are defined in eqs.~\eqref{susyb20}. The last expression, a standard Born--Infeld Lagrangian with a $\vartheta$--term, is recovered integrating out the auxiliary field ${\cal F}$.

In the multi--field case the canonically normalized Goldstino superfield is
\beq
W_{G\,\a} \ = \ \left( \frac{i}{2}\ \bar{Q}^T \, \Omega\, Q \right)^{\,-\,\frac{1}{2}} e_{2\,\Lambda}\ W_\a^\Lambda \ .
\eeq{4aaa2}
Its non--linear variation under the second supersymmetry is
\beq
\delta \, W_{G\,\a} \ = \ \left( \frac{i}{2}\ \bar{Q}^T \, \Omega\, Q \right)^{\,-\,\frac{1}{2}} e_{2\,\Lambda}\ m^\Lambda \ \eta_\a \ + \ \ldots \ = \ \left( \frac{i}{2}\ \bar{Q}^T \, \Omega\, Q \right)^{\,\frac{1}{2}} \ \eta_\a \ + \ \ldots \ ,
\eeq{4ab}
so that in units of $M$ the supersymmetry breaking scale is
\beq
f^\frac{1}{2} \ = \ \left( \frac{i}{2}\ \bar{Q}^T \, \Omega\ Q \right)^{\,\frac{1}{4}} \ ,
\eeq{4abc}
according to eqs.~\eqref{susyb9} and \eqref{susyb12}.

%%%%%%%%%%%%%%%%%%%%%%%%%%%%%%%%%%%%%%%%%%%%%%%%%%%%%%%
\scs{Analogies with $N=2$ extremal black holes}\label{sec:5}
%%%%%%%%%%%%%%%%%%%%%%%%%%%%%%%%%%%%%%%%%%%%%%%%%%%%%%%

The preceding discussion presents some analogies with the attractor equations for $N=2$ extremal black holes with symplectic
vector $Q=(m^\Lambda,e_\Lambda)$, which we would like to stress. Indeed, the black hole potential~\cite{attractors},
\beq
V_{BH}\ =\ \frac{1}{2} \ Q^T {\cal M} \, Q \ ,
\eeq{24bb}
is also determined in terms of the matrix ${\cal M}$ of eq.~\eqref{17c}, via the last expression in eq.~(\ref{21}), albeit for a real $Q$, so that the $\Omega$ term vanishes identically. Moreover, in this case the value attained by $V_{BH}$ at the attractor point is positive and gives the Bekenstein--Hawking entropy
\beq
V_{BH}(X_{attr})\ = \ \frac{A}{4\, \pi} \ =\ \frac{S(Q)}{\pi} \ .
\eeq{25}
When expressed in terms of the central charge $Z$, which is the counterpart of ${\cal W}$, the black--hole potential contains an additional term~\cite{attractors}, and reads \footnote{Note that the second term in \eqref{26} is a consequence of local Special Geometry \cite{stro}.}
\beq
V_{BH} \ = \ |{\cal D}_i Z|^2 \ + \ |Z|^2\ .
\eeq{26}
Hence, at the $\frac{1}{2}$ -- BPS critical point, where ${\cal D}_i \, Z\,=\,0$,
\beq
V_{crit} \ \equiv \ V_{BH}(X_{attr}) \ = \ |Z|^2_{attr} \ .
\eeq{27}
On the other hand, in our case $V_{crit}=0$, which implies $\frac{\partial \, {\cal W}}{\partial X^\Lambda}=0$ in order to leave $N=1$ supersymmetry unbroken.

%%%%%%%%%%%%%%%%%%%%%%%%%%%%%%%%%%%%%%%%%%%%%%%%%%%%%%%
\scs{$N=2$ Born--Infeld attractors and cubic invariants}\label{sec:6}
%%%%%%%%%%%%%%%%%%%%%%%%%%%%%%%%%%%%%%%%%%%%%%%%%%%%%%%

Our goal in this section is to discuss solutions of the quadratic constraints of eq.~\eqref{30d}, which we rewrite here for convenience
\beq
d_{\Lambda\Sigma\Gamma} \left[ F_+^\Sigma\cdot F_+^\Gamma \ + \ {\cal F}^\Sigma \left( m^\Gamma \ - \ \bar{\cal F}^\Gamma\right)\right] \ = \ 0 \ .
\eeq{abc2}
To this end, it is convenient to consider explicitly the real and imaginary parts of these constraints, while also letting
\beq
{\cal H}^\Sigma \ = \ \frac{m^\Sigma}{2} \ - \ Re \,{\cal F}^\Sigma \ , \qquad {\cal R}^{\Sigma\Gamma} \ = \ F^\Sigma\cdot F^\Gamma \ + \frac{m^\Sigma\,m^\Gamma}{4} \ - \ Im\, {\cal F}^\Sigma \, Im\, {\cal F}^\Gamma \ .
\eeq{4}
The real parts of eq.~\eqref{abc2} thus yield the system of coupled quadratic equations
\beq
d_{\Lambda\Sigma\Gamma} \left( {\cal H}^\Sigma\, {\cal H}^\Gamma \ - \ {\cal R}^{\Sigma\Gamma} \right) \ = \ 0 \ ,
\eeq{5}
while their imaginary parts are $n$ linear equations for the $Im\, {\cal F}^\Lambda$:
\beq
d_{\Lambda\Sigma\Gamma} \left( F^\Sigma \cdot {\widetilde{F}}^\Gamma \ + \ m^\Sigma \, Im\, {\cal F}^{\,\Gamma}  \right) \ = \ 0 \ .
\eeq{6}

Any specific class of models with $n$ vector multiplets, obtained solving the constraints of eqs.~\eqref{abc2}, is defined via the polynomials
\beq
U \ = \ \frac{1}{3!} \ d_{\Lambda\Sigma\Gamma}\,X^\Lambda\,X^\Sigma\, X^\Gamma \ ,
\eeq{bia1}
modulo field redefinitions by projective $SL(n,R)$ transformations. Inequivalent theories are thus classified by the orbits $P(Sym^3 (R^n))$ of the three--fold symmetric product of the fundamental representation of $SL(n,R)$. Distinct types of polynomials are classified according to different degenerations of the cubic $(n-2)$--projective varieties defined by $U=0$. These varieties are points for $n=2$, curves for $n=3$, surfaces for $n=4$, and so on: their classification problem was completely solved by Mathematicians for $n=2,3$, while only partial results are available for $n \geq 4$ \cite{inv_theory}. These results allowed to define completely, in \cite{fps}, the generalized Born--Infeld models induced by the $N=2$ Special Geometry for the cases of two and three vector multiplets. For brevity, here we shall display the complete results over the real numbers only for $n=2$, while for $n=3$ we shall content ourselves with a brief review of the simpler classification over the complex numbers. The richer structure that emerges over the reals for $n=3$ is described in detail in \cite{fps}.

In the $n=2$ case, the relevant representation corresponding to the $d_{ABC}$ is the spin--$\frac{3}{2}$ of $SL(2,R)$ \footnote{It is actually a special case of the famous Cayley hyperdeterminant, which also plays a role for $q$--bit entanglement in Quantum Information Theory.}. It is known to possess a unique quartic invariant that actually corresponds to the discriminant $I_4$ of the polynomial
\beq
I_4\ =\ -\ 27 \,d_{222}^{\,2} \,d_{111}^{\,2} \ + \ d_{221}^{\,2} \, d_{112}^{\,2} \ + \ 18\, d_{222}\,d_{111}\, d_{112}\, d_{221} \ -\ 4 \, d_{111}\ d_{122}^{\,3}
\ -\ 4\, d_{222}\, d_{211}^{\,3}\ .
\eeq{bia11}
Its orbits were classified, and simply correspond to the possible degenerations of the three roots of a cubic equation. Different configurations of the roots are associated to different properties of its four orbits: $O_{t},O_{s}, O_{l}, O_{c}$.

For $I_4>0$ the cubic has three real simple roots and $O_{t}$ is a \emph{``time--like''} orbit. When the roots are simple but two
are complex conjugates, $I_4<0$ and the orbit $O_{s}$ is \emph{``space--like''}. A double root $I_4=0$, $\partial I_4 \neq 0$
corresponds to a \emph{``light--like''} orbit $O_{l}$, and finally a triple root corresponds to $I_4=\partial I_4=0$ and to the \emph{``critical''}
orbit $O_{c}$ comprises a single point. This language is drawn from the black--hole literature, for which we refer to \cite{ferborst}.

The four inequivalent theories can be associated to the four representative polynomials determined by the conditions
\beq
\begin{array}{lll}
I_4 \, >\, 0 \qquad & d_{222}\, =\, d_{211}\, \neq \, 0  \qquad & O_{t} \ , \\
I_4 \, < \, 0 &  d_{222}\, =\, d_{111}\, \neq \, 0 & O_{s} \ , \\
I_4\, =\, 0 & d_{222}\, =\, d_{221}\, \neq \, 0 & O_{l} \ ,\\
\partial \, I_4 \, =\, 0 & d_{222}\, \neq \, 0 & O_{c} \ ,
\end{array}
\eeq{45}
which read
\bea
&& O_t \ = \ \frac{1}{3!} \ X^{\,3} \ - \ \frac{1}{2} \ X \, Y^{\,2} \ , \\
&& O_s \ = \ \frac{1}{3!} \left(  X^{\,3} \ + Y^{\,3} \right) \ , \\
&& O_l \ = \ \frac{1}{3!} \ X^{\,3} \ - \ \frac{1}{2} \ X^{\,2} \, Y \ , \\
&& O_c \ = \ \frac{1}{3!} \ X^{\,3} \ .
\eea{45a}
The imaginary parts of the Hessian matrices of these polynomials contribute to the kinetic terms. It is simple to see that only in the $O_s$ case the Hessian is positive definite. On the other hand, the Hessians of the $O_t$ and $O_l$ cases have negative determinant, so that their eigenvalues have opposite signs. Finally, in the $O_c$ case there is a vanishing eigenvalue. Hence, aside from the $O_s$ case a $C_{AB}$ term is needed in the generalized BI Lagrangians. The explicit solutions of the constraints \eqref{abc2} are relatively simple to find in this case, and are illustrated in \cite{fps}. Their typical feature is the presence of double radicals.

On the other hand, in the $n=3$ case the discriminant of the cubic \eqref{bia1} is of degree 12, and is built from the particular combination
\beq
I_{12} \ = \ P_4^3 \ - \ 6\, Q_6^2 \ ,
\eeq{bia2}
where the two invariants of the ten--dimensional $Sym^3 (R^3)$, of order four and six, are
\begin{eqnarray}
P_4&=&d_{a_1\,a_2\,a_3}\,d_{b_1\,b_2\,b_3}\,d_{c_1\,c_2\,c_3}\,d_{d_1\,d_2\,d_3}\epsilon^{b_1\, a_1\, d_1}\epsilon^{c_2\, d_2\, a_2}
\epsilon^{b_3\, c_3\, a_3}\epsilon^{d_3\, c_1\, b_2}\, ,\\
Q_6&=&d_{a_1\,a_2\,a_3}\,d_{b_1\,b_2\,b_3}\,d_{c_1\,c_2\,c_3}\,d_{d_1\,d_2\,d_3}\,d_{f_1\,f_2\,f_3}\,d_{h_1\,h_2\,h_3}\epsilon^{h_3\, a_1\, b_1}\epsilon^{f_3\, c_1\, a_2}
\epsilon^{d_3\, b_2\, c_2}\epsilon^{c_3\, f_2\, d_2}\epsilon^{a_3\, h_2\, f_1}\epsilon^{b_3\, d_1\, h_1}\, .\nonumber
\end{eqnarray}
\begin{figure}[h]
\begin{center}
\epsfig{file=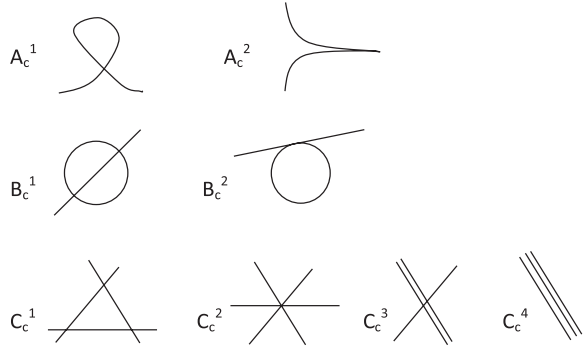, height=1.8in, width=3in}
\end{center}
\caption{\small
Different types of degenerations for complex curves with $I_{12}=0$.
}
\label{fig:sing_curves}
\end{figure}

In this case the inequivalent types of orbits reflect the different degenerations of cubic curves. To begin with, for non--degenerate curves $I_{12} \neq 0$, while for degenerate ones it vanishes. A finer classification of these cases is captured by the orders of $I_{12}$, $P_4$ and $Q_6$, whose derivatives can vanish to different orders. All these cases fall in two groups, distinguished by the pair of values of $P_4$ and $Q_6$. When these do not vanish, $\partial I_{12}$ and $\partial^{\,2}I_{12}$ may or may not vanish. Moreover, when $P_4$ and $Q_6$ both vanish, different derivatives of these invariant polynomials may or may not vanish. In detail, the correspondence is as follows:
\begin{itemize}
\item  $A_C^1 \leftrightarrow (P_4,Q_6) \neq (0,0)$; \ $A_C^2 \leftrightarrow \partial I_{12}=0, (P_4,Q_6) = (0,0)$;
\item $B_C^1 \leftrightarrow \partial I_{12}=0, (P_4,Q_6) \neq (0,0)$;  \ $B_C^2 \leftrightarrow \partial^2 I_{12}=0, (P_4,Q_6) = (0,0), \partial Q_6 = 0$;
\item $C_C^1 \leftrightarrow \partial^2 I_{12}=0, (P_4,Q_6) \neq (0,0)$; \ $C_C^2 \leftrightarrow \partial^2 I_{12}=0, (\partial P_4, \partial Q_6) = (0,0)$;
\item $C_C^3 \leftrightarrow \partial^2 I_{12}=0, \partial^2 Q_6 = 0$; $C_C^4 \leftrightarrow \partial^2 I_{12}=0, (\partial^2 P_4, \partial^2 Q_6) = (0,0)$;
\end{itemize}
For each of these models there is a representative normal form, which allows a systematic exploration of different realizations of special geometry  with three vector multiplets. The Fayet--Iliopoulos terms of the $N=2$ theory are in this case $SU(2)$ triplets of $Sp(6,R)$ charge vectors \cite{fps,adt_15,adft_15}. Remarkably, the non--linear systems of eqs.~\eqref{abc2}, which are now rather involved, are exactly solvable in terms of radicals in all these cases (see the Appendix of the second paper in \cite{fps}).

The orbit classification in the $n=2$ systems is curiously connected to the extremal black--hole classification of the $N=2$ Supergravity $T^3$ model \cite{cdfy}, in which the electric and magnetic charges of the four spin--$\frac{3}{2}$ fields correspond to the electric and magnetic charges of the graviphoton and the matter photon. Therefore, there is a dictionary converting types of extremal black--hole in this model to the classification of these rigid $n=2$ theories.

%%%%%%%%%%%%%%%%%%%%%%%%%%%%%%%%%%%%%%%%%%%%%%%%%%%%%%%
\scs{Nilpotent superfields in curved Superspace and Cosmology }\label{sec:7}
%%%%%%%%%%%%%%%%%%%%%%%%%%%%%%%%%%%%%%%%%%%%%%%%%%%%%%%

Nilpotency constraints can be easily implemented in local Supersymmetry, since their definition can be extended to curved Superspace if the constraints afford an off--shell description. This is clearly the case for $N=1$ Supersymmetry, and also for $N=2$ when only vector multiplets are present. For example, pure Supergravity coupled to the goldstino (Volkov--Akulov superfield) $S$, such that $S^2 = 0$, allows a two--parameter Superpotential
\beq
W \ = \ f\,S \ + \ W_0 \ ,
\eeq{sugra1}
which describes the consistent theory of a massive (massless) gravitino coupled to gravity, in the presence of the cosmological constant
\beq
\Lambda \ = \ \left|f\right|^2 \ - \ 3\, \left| W_0 \right|^2 \ .
\eeq{sugra2}
Supersymmetry is broken if $f \neq 0$, so that unbroken Supersymmetry requires maximally symmetric spaces that are Minkowski or AdS according to whether $W_0$ does or does not vanish. On the other hand, if $f \neq 0$ also dS is allowed, corresponding to the inflationary regime in Cosmology \cite{dz,superhiggs_1}.

The minimal extension of this model is the Volkov--Akulov--Starobinsky Supergravity, which encodes the interactions of the Volkov--Akulov non--linear multiplet with a chiral superfield that contains the inflaton, with Kahler potential
\beq
K \ = \ -\ 3 \, \log \left( T \, +\, \overline{T} \, - \, S\, \overline{S} \right)
\eeq{sugra3}
and superpotential
\beq
W \ = \ M\,S\,T \ + \ f\, S \ + \ W_0 \ .
\eeq{sugra4}
After the redefinition
\beq
T \ = \ e^{\phi \, \sqrt{\frac{2}{3}}} \ + \ i\, a\, \sqrt{\frac{2}{3}} \ ,
\eeq{sugra5}
which brings the scalar kinetic terms in canonical form, the Lagrangian takes the form \cite{adfs}
\beq
{\cal L} \ = \ \frac{R}{2} \ - \ \frac{1}{2}\ (\partial \phi)^2 \ - \ \frac{1}{2} \ e^{\,-\,2\, \phi \,{\sqrt \frac{2}{3}}}\
(\partial a)^2 \ - \ \frac{M^{\,2}}{12} \left(1 \ - \ e^{\,-\,{\sqrt \frac{2}{3}} \ \phi}\right)^2 - \frac{M^2}{18} \ e^{- 2\,\phi
{\sqrt \frac{2}{3}}} \ a^2 \, ,
\eeq{sugra6}
after eliminating $f$ by a shift of $\phi$ and a constant rescaling of $a$. This is precisely a minimal Starobinsky Lagrangian where the inflaton $\phi$ is accompanied by an axion field $a$, which is however far heavier during the inflationary phase, since
\beq
m_{\phi}^2 \ \simeq \ \frac{M^2}{9} \ e^{\,-\,2\, \phi_0 {\sqrt \frac{2}{3}}} \ \ll \
m_{a}^2 \ \equiv \ \frac{M^2}{9} \ .
\eeq{sugra7}
Note that the potential in eq.~\eqref{sugra6} is non negative due to the no--scale structure of the Kahler potential \cite{noscale,noscale_2,noscale_3}. Nilpotent superfields were previously considered in a different context, still related to inflation, in \cite{lag}.

The coupling of the Volkov--Akulov model to Supergravity and to the inflaton was generalized to several cases which all share the property that the ``sgoldstino'' multiplet is identified with the Volkov--Akulov multiplet \cite{fkl_1}. The results are collectively referred to as ``sgoldstino--less models'', and include some interesting recent constructions that incorporate separate scales of Supersymmetry breaking, during and at the exit of inflation \cite{fkl_2,dallz,fkl_3}. All these models have a trivial Kahler geometry, with
\beq
K(\Phi,S) \ =  \ \frac{1}{2} \left( \Phi \ + \ \overline{\Phi}\right)^2 \ + \ S\, \overline{S} \ ,
\eeq{sugra8}
where
\beq
\Phi \ = \ \frac{1}{\sqrt{2}} \left( a  \ + \ i\, \varphi \right) \ ,
\eeq{sugra55}
so that the shift symmetry present in $K$ protects these systems from the $\eta$--problem.
One can distinguish two classes of models of this type, which differ in the supersymmetry breaking patterns during and after inflation.

In the first class of models \cite{fkl_2}
\beq
W(\Phi,S) \ = \ M^2 \,S\left[ 1\ + \ g^2(\Phi) \right] \ + \ W_0 \ ,
\eeq{sugra66}
where $g$ vanishes at the origin. Along the inflaton trajectory, $Re(\Phi)=0$, the potential takes the form
\beq
V \ = \ M^4 \left| g(\varphi)\right|^2 \left[ 2 \ + \ |g(\varphi)|^2 \right] \ + \ V_0 \ ,
\eeq{sugra77}
where
\beq
V_0 \ = \ M^4 \ - \ 3\, W_0^2 \ .
\eeq{sugra88}
If $V_0 \simeq 0$ inflation ends in Minkowski space, and recalling that $M^4 = H^2 \, M_P^2$ one finds
\beq
m_\frac{3}{2} \ = \ \frac{1}{\sqrt{3}} \ H \ , \quad E_{SB} \ = \ \left| F_S \right|^\frac{1}{2} \ = \ \sqrt{H\, M_P} > H \ ,
\eeq{sugra9}
for the gravitino mass and the supersymmetry breaking scale.
Note that during the inflationary phase, where $Re(\Phi)=0$, $F_\Phi$ also vanishes, so that inflaton potential reduces to
\beq
V \ = \ F_S\, F^S \ - \ 3\, \left| W_0 \right|^2 \ .
\eeq{sugra10}

In the second class of models \cite{dallz}
\beq
W(\Phi,S) \ = \ f(\Phi) \left[ 1 \ + \ \sqrt{3}\, S \right] \ ,
\eeq{sugra11}
with
\beq
\overline{f}(\Phi)  \ = \ f(-\overline{\Phi}) \ , \quad f^\prime(0) \ = \ 0 \ , \quad f(0) \ \neq \ 0 \ .
\eeq{sugra12}
The resulting scalar potential is of no--scale type \cite{noscale,noscale_3}, since
\beq
F^S\,F_S \ = \ 3\, e^G \ = \ 3\, e^{a^2} \, \left| f(\Phi) \right|^2
\eeq{sugra13}
and then
\beq
V(a,\varphi) \ = \ F^\Phi\,F_\Phi\ = \ e^{a^2} \left| f^\prime(\Phi) \ + \ a\, \sqrt{2} \, f(\Phi) \right|^2 \ .
\eeq{sugra14}
Notice that $a$ is stabilized at the origin since $f$ is even in $a$. During inflation $a$ acquires a mass of order $H$ without mixing with $\Phi$ and is rapidly driven to zero. As a result, the inflationary potential is
\beq
V(a=0,\varphi) \ = \ \left| f^\prime\left( \frac{i\varphi}{\sqrt{2}} \right)\right|^2 \ ,
\eeq{sugra15}
and vanishes at the origin on account of eq.~\eqref{sugra12}.

It is interesting to compare the supersymmetry breaking patterns. In these models $F^\Phi$ does not vanish during inflation, while and at the end of inflation
\beq
\langle F^\Phi \rangle \ = \ 0 \ , \quad \langle F^S \rangle \ = \ \sqrt{3}\ \overline{f}(0) \ , \quad m_{\frac{3}{2}} \ = \ \left|f(0)\right| \ ,
\eeq{sugra17}
so that
\beq
F^S\,F_S \ = \ 3\, e^{G(0,0)} \ = \ 3\, m_{\frac{3}{2}}^2 \ .
\eeq{sugra16}
For instance, a choice that reproduces the Starobinsky potential is
\beq
f(\Phi) \ = \ \lambda \ - \ i\, \mu_1 \, \Phi \ + \ \mu_2 \, e^{i\,\frac{2}{\sqrt{3}}\,\Phi} \ ,
\eeq{sugra177}
with $\mu_1$ and $\mu_2$ real constants. Interestingly, $m_a$ and $m_{\frac{3}{2}}$ depend on the integration constant $\lambda$ but $m_\varphi$ does not, since $V$ is independent of $\lambda$.

Non--linear realizations play a central role in some vacua of String theory where supersymmetry is broken at the scale of string excitations. The simplest realization of this phenomenon, usually referred to as ``brane supersymmetry breaking'', is in the ten--dimensional Sugimoto model in \cite{BSB}, where the uncanceled tensions of branes and orientifolds manifest themselves in the string--frame dilaton potential
\beq
V \ = \ V_0 \, e^{-\phi} \ .
\eeq{sugra18}
As discussed in \cite{dm}, this term brings along a non--linear realization of Supersymmetry in ten dimensions, while this type of potential is ``critical'', in the sense of \cite{dks}, since in the corresponding flat cosmologies the dilaton is barely ``forced to climb it up'' as it emerges from an initial singularity. Moreover, a ``critical'' potential also builds up, for $D<10$, for dimension--dependent combinations of dilaton and breathing mode, so that this type of behavior is typical for compactifications of these types of systems \cite{mixing}.

In lower dimensions this term mixes with contributions from the breathing mode, giving rise again, for all $D<10$, to a critical exponent. In four dimensions, in the Einstein frame and for a canonically normalized scalar $\phi$, the corresponding potential would read
\beq
V \ = \ V_0 \, e^{\,\pm \,\sqrt{6}\,\phi} \ .
\eeq{sugra19}
Following \cite{fkl_1}, we can simply discuss how potentials of this type can be recovered working with a nilpotent superfield. This is also interesting since they provide the rationale for the KKLT \cite{kklt} $F$--term uplift, as first stressed in \cite{dks}.

The idea is simply to turn the Kahler potential of eq.~\eqref{sugra3} into
\beq
K \ = \ -\ 3 \, \log \left( T \, +\, \overline{T}  \right) \ + \ \frac{S\,\overline{S}}{\left(T+\overline{T}\right)^\alpha}\ ,
\eeq{sugra20}
and for our present purpose the superpotential
\beq
W \ = \ W_0 \ + \ f\, S
\eeq{sugra21}
will suffice. The resulting potential, obtained as before setting eventually to zero $S$, which contains no independent scalar mode, reads
\beq
V \ = \ |f|^2 \ \left( T+ \overline{T}\right)^{\alpha -3} \ .
\eeq{sugra22}
After the redefinition \eqref{sugra5} its dependence on $\phi$
thus takes the form
\beq
V \ = \ 2^{\,\alpha-3}\ |f|^2 \ e^{\,(\alpha -3) \, \phi\, \sqrt{\frac{2}{3}}} \ ,
\eeq{sugra222}
so that the critical potential of \cite{dks} obtains for $\alpha=0$, while an opposite ``critical'' exponent obtains for $\alpha=6$. The former case was discussed in detail, in this context, in \cite{fkl_2,fkl_3}. The resulting axion--dilaton system has the interesting property of only allowing a climbing behavior close to the initial singularity, despite the presence of a flat axion potential \cite{dks}.

%%%%%%%%%%%%%%%%%%%%%%%%%%%%%%%%%%%%%%%%%%%%%%%%%%%%%%%
\scs{Dual non--linear systems of higher forms}\label{sec:8}
%%%%%%%%%%%%%%%%%%%%%%%%%%%%%%%%%%%%%%%%%%%%%%%%%%%%%%%

%%%%%%%%%%%%%%%%%%%%%%%%%%%%%%%%%%%%%%%%%%%%%%%%%%%%%%%
\scss{Massless dualities}\label{sec:8.1}
%%%%%%%%%%%%%%%%%%%%%%%%%%%%%%%%%%%%%%%%%%%%%%%%%%%%%%%

The partial breaking of extended supersymmetry is closely linked to the physics of branes, as was originally observed in \cite{partial_n2_3}.
Different realizations of $N=2$ Supersymmetry spontaneously broken to $N=1$ give rise to physically different non--linear Lagrangians whose propagating degrees of freedom are an $N=1$ vector multiplet \cite{partial_n2_6}, or alternatively an $N=1$ tensor multiplet or chiral multiplet, where the latter two options are dual to one another \cite{partial_n2_7}. For instance, the Supersymmetric Born--Infeld action \cite{borninfeldsusy_1,borninfeldsusy_2} inherits from its bosonic counterpart, which is the standard Born--Infeld action \cite{borninfeld}, its self--duality. However, the tensor and chiral multiplet actions enjoy a different type of duality, where an antisymmetric tensor is turned into a scalar and/or vice versa. This new duality, which we call ``double self--duality'', leads to three dual Lagrangians, depending on whether the two spinless massless degrees of freedom are described via a scalar and antisymmetric tensor, two scalars or two antisymmetric tensors. While in the first case the action turns out to be doubly self--dual \cite{partial_n2_7}, in the other cases a double duality maps one action into the other. The three actions are also connected by a single duality affecting only one of the two fields.

As pointed out in \cite{partial_n2_3,partial_n2_8}, in four dimensions the close connection between the Supersymmetric Born--Infeld action and the non--linear tensor multiplet action stems from similarities between the superspace $N=1$ constraints underlying the two models. Indeed, introducing the $N=1$ vector multiplet chiral field strength $W_\alpha=\overline{D}^{\,2} \, D_\alpha V$ ($\overline{D}_{\dot{\alpha}}W_\alpha=0$) and the corresponding object $\psi_\alpha = D_\alpha L$ ($D_\alpha \, \psi_\beta=0$) for a linear multiplet ($D^{\,2}L = {\overline{D}}^{\,2}L=0$), the non--linear actions in the two cases are determined by the non--linear constraints
\beq
X \, = \, - \ \frac{W_\alpha^{\,2}}{\mu \ - \ {\overline{D}}^{\,2} \,\overline{X}}  \qquad {\rm and } \qquad X \, = \, - \ \frac{\psi_\alpha^{\,2}}{\mu \ - \ {D}^{\,2} \,\overline{X}}\ ,
\eeq{1}
where $\mu$ is a parameter with mass--square dimension that sets the supersymmetry breaking scale. $X$ is chiral in the first case, since $W_\alpha$ is a chiral superfield, and is antichiral in the second, since $\psi_\alpha$ is an antichiral one. However, the highest components of the equations are identical, provided one maps the complex field
\beq
G_+^{\,2} \ = \ F^{\,2} \ + \ i\, F\, \widetilde{F}
\eeq{2}
of the Born--Infeld action into the complex field
\beq
\left( \partial \phi \right)^2 \ - \ H_{\mu}^{\,2}  \ + \ i\, H^{\mu}\, \partial_\mu \phi
\eeq{3}
of the linear multiplet action, where
\beq
H_\mu \ = \ \frac{1}{3!} \ \epsilon_{\mu\nu\rho\sigma}\, \partial^\nu\, B^{\rho\sigma} \ .
\eeq{4444}
For both systems, the non--linear Lagrangian is proportional to the $F$--component of the chiral(antichiral) superfield $X$. This is subject to the constraint in \eqref{1}, which implies in both cases the nilpotency condition $X^2=0$.

In this section we briefly review some of the results in \cite{fsy}, which generalize the setup to pairs of forms in $D$ dimensions with complementary field strengths
\beq
H_{p+1} \ = \ d\, B_p \, \qquad V_{D-p-1} \ = \ d\, A_{D-p-2}  \ ,
\eeq{5555}
so that one can write the geometrical term
\beq
H_{p+1} \wedge V_{d-p-1} \ .
\eeq{6666}
General duality properties for massless higher--form gauge fields, including some of the models considered here, were previously studied by in \cite{kt}.

The tensor multiplet non--linear Lagrangian enjoys a \emph{double self--duality}, in the following sense. To begin with, one can either dualize the scalar into a two--form or the two--form into a scalar. In both cases, the resulting Lagrangian involves two fields of the same type and is symmetric under their interchange. Hence, double self--duality is guaranteed by the symmetry, since two successive Legendre transforms yield the identity. This result therefore applies to the class of non--linear Lagrangians related to the Born--Infeld one and brought about by Supersymmetry, but also, in principle, to more general ones. All these systems can be formulated in terms of a pair of complex Lagrange multipliers \cite{partial_n2_8,abmz}: the first provides a non--linear constraint, whose solution determines the value of the second, which in its turn determines the non--linear action in a square--root form. The pattern is along the lines of what was discussed in detail in \cite{fps,fs15,fsy}.

The supersymmetric Born--Infeld action inherits a key property of its bosonic counterpart, in that it is self--dual in Superspace. One can explore the mass generation mechanism that the addition of some geometric coupling can induce in these non--linear theories. It is connected to the Stueckelberg mass generation mechanism for $p$--form gauge fields, which afford a dual formulation in terms of $(D-p-1)$--form gauge fields, where the dual mass term originates from a Green-Schwarz coupling of their $(D-p)$--form field strength to a $p$--form gauge field, or equivalently of their $(D-p-1)$--form gauge field to a $p+1$--form field strength. As a result, the non--linear curvature interactions present in one formulations are dual to non--linear generalizations of the mass term present in the other. Massive generalizations of doubly self--dual systems are also possible, as discussed in \cite{fsy}.

%%%%%%%%%%%%%%%%%%%%%%%%%%%%%%%%%%%%%%%%%%%%%%%%%%%%%%%
\scss{Massive supersymmetric Born--Infeld and its dual}\label{sec:8.2}
%%%%%%%%%%%%%%%%%%%%%%%%%%%%%%%%%%%%%%%%%%%%%%%%%%%%%%%

In four dimensions a massive vector is dual to a massive antisymmetric tensor, and the Stueckelberg mechanism for the former finds a counterpart in an ``anti--Higgs'' mechanism for the latter (in the sense that a massless $B_{\mu\nu}$, with one degree of freedom, eats a massless vector carrying two degrees of freedom).

This relation becomes particularly interesting in the supersymmetric context since, as we have stressed in the preceding sections, the supersymmetric extension of the Born--Infeld action is the Goldstone action for $N=2$ spontaneously broken to $N=1$. It thus describes the self interactions of an $N=1$ vector multiplet whose fermionic component, the gaugino, plays the role of Goldstone fermion for the broken Supersymmetry. When coupled to Supergravity, this system is expected to be a key ingredient in models for the $N=2 \to N=1$ super-Higgs effect of partially broken Supersymmetry. Consequently, the gaugino must be eaten by the gravitino of the broken Supersymmetry, which then becomes massive. In fact, because of the residual Supersymmetry the massive gravitino must complete an $N=1$ massive multiplet, which also contains two vectors and a spin--$\frac{1}{2}$ fermion \cite{FVN}. The Born--Infeld system, which contains in the rigid case a massless vector as partner of the Goldstone fermion, must therefore become massive. This provides in general a motivation \cite{fs15} to address massive Born--Infeld systems for $p$--forms \cite{cfg}.

In four--dimensional $N=1$ superspace the vector $A$ belongs to a real superfield $V$, while the two-form $B$ belongs to a spinor chiral multiplet $L_\alpha$ ($ \overline{D}_{\dot{\alpha}} \, L_\alpha = 0$). The two--form field strength $dA$ belongs to the chiral multiplet
\beq
W_\alpha(V) \ = \ \overline{D}^{\,2} D_\alpha V \ ,
\eeq{233}
which is invariant under the superspace gauge transformation
\beq
V \ \rightarrow \ V \ + \ \Lambda \ + \overline{\Lambda}  \qquad (\overline{D}_{\dot{\alpha}} \, \Lambda \ = \ 0) \ .
\eeq{23a}
This is the superspace counterpart of the familiar Maxwell gauge transformation $A \to A \ + \ d \lambda$. On the other hand, the three--form field strength $H=dB$ belongs to a linear multiplet $L$, which is related to $L_\alpha$ according to
\beq
L \ = \ i \left( D^\alpha\, L_\alpha \ - \ \overline{D}_{\dot{\alpha}}\, \overline{L}^{\,\dot{\alpha}}\right) \ ,
\eeq{24cc}
and satisfies the two constraints
\beq
D^2 L \ = \ \overline{D}^2 L \ = \ 0 \ .
\eeq{24aa}
Note that $L$ is invariant under the gauge transformation
\beq
L_\alpha \ \rightarrow \ L_\alpha \ + \ W_\alpha(Z)\ ,
\eeq{25aaa}
for any real superfield $Z$, due to the superspace identity
$D^\alpha \overline{D}^{\,2} D_\alpha = \overline{D}_{\dot{\alpha}} D^{\,2} \overline{D}^{\dot{\alpha}}$. Eq.~\eqref{25} is the superspace counterpart of the gauge transformation $B \to B \ + \ d z$, where $z$ is a one--form~\footnote{Strictly speaking, the linear multiplet $L$ is the super field strength of the chiral multiplet $L_\alpha$. Only the latter ought to be called tensor multiplet, because it contains the tensor field $B_{\mu\nu}$. With a slight abuse of language, however, we use loosely the term linear multiplet for both.}.

In superfield language, the supersymmetric Born--Infeld action takes the form
\beq
{\cal L}_{BI}  \ = \  \left.{\Xi}\big[g\, , \,\mu\, , \, W^2(V)\, , \,\overline{W}^{\,2}(V)\big]\right|_D \  + \ \left. \frac{1}{2\, g^{\,2}} \ \left( W^2(V) \ + \ {\rm h.c.} \right) \right|_F\ ,
\eeq{26asa}
where ${\Xi}$ is given in \cite{borninfeldsusy_2,partial_n2_6}. This expression clearly reduces to the supersymmetric Maxwell action for ${\Xi}=0$, and the leading non--linear order correction is clearly proportional to $\left. \frac{1}{\mu^{\,2}\,g^{\,2}}\ W^2(V)\,\overline{W}^{\,2}(V) \right|_D$.

It is now convenient to recast eq.~\eqref{26} in the first--order form
\beq
{\cal L}_{BI} \ = \ \left.\phantom{\frac{1}{2}} {\Xi}\big[g\,,\,\mu\, ,\, W^2,\overline{W}^{\,2}\big]\right|_D \ + \ \left. \frac{1}{2\,g^{\,2}} \ \left( W^2 \ + \ {\rm h.c.} \right) \right|_F
\ +  \left. \phantom{\frac{1}{2}} i \left( M^\alpha \, W_\alpha \ + \ h.c. \right)\right|_F
\ ,
\eeq{27aaa}
introducing a dual potential $V_D$ and letting
\beq
M_\alpha = \overline{D}^{\,2}\, D_\alpha V_D \ .
\eeq{27aba}
Integrating out $V_D$ one recovers the original action \eqref{26}, while integrating out $W_\alpha$ yields the condition
\beq
\overline{D}^{\,2}\ \frac{\partial {\cal L}_{BI}}{\partial \, W^\alpha} \ \equiv \ \overline{D}^{\,2}\, \frac{\partial \,{\Xi}}{\partial \,W^\alpha} \ + \ \frac{1}{g^{\,2}} \ W_\alpha \ = \ - \ i \ M_\alpha \ .
 \eeq{28}
The equation of motion then follows from the Bianchi identity of the dual field strength $W_\alpha(V_D)$,
\beq
D^\alpha\, M_\alpha \ - \ \overline{D}_{\dot{\alpha}}\, \overline{M}^{\dot{\alpha}} \ = \ 0 \ .
\eeq{29aaa}
As discussed in \cite{partial_n2_6,kt}, the superspace Born--Infeld action enjoys a self--duality, which is the direct counterpart of what we have seen in components in eq.~\eqref{17} and translates into the condition \cite{kt}
\beq
\left. \phantom{\frac{1}{2}} Im \left[ \ M^\alpha \, M_\alpha \ + \ \frac{1}{g^{\,4}} \ W^\alpha \, W_\alpha \ \right]\right|_F \ = \ 0 \ ,
\eeq{30}
where $Im$ picks the imaginary part of the $F$--component.
Notice that this is trivially satisfied in $N=1$ Maxwell Electrodynamics, on account of the special form of eq.~\eqref{28} when ${\Xi}$ vanishes.

We can now turn to the supersymmetric version of the massive duality between one--form and two--form gauge fields in four dimension. The corresponding master Lagrangian, written in first--order form,
\beq
{\cal L} \ = \ \Phi(U) \ + \ L(U \ - \ m V) \ + \ {\cal L}_{BI} \left[g\,,\,\mu\, ,\, W_\alpha(V),  \overline{W}_{\dot{\alpha}}(V)\right] \ ,
\eeq{31}
where $L$ is a linear multiplet Lagrange multiplier and $U$ and $V$ are real superfields, encompasses both formulations. The Lagrangian \eqref{31}, with the last term replaced by a standard quadratic super--Maxwell term $W^\alpha W_\alpha$, was considered in connection with $R+R^2$ theories in \cite{CFPS} and, more recently, for models of inflation, in \cite{FKLP}. $U$ is the superfield extension of an unconstrained $H_{\mu\nu\rho}$, and the presence of the arbitrary function $\Phi(U)$ reflects the freedom to dress the tensor kinetic term with an arbitrary function of the scalar field present in the linear multiplet.

Varying the action with respect to $L$ gives
\beq
U \ - \ m\, V \ = \ T \ + \ \overline{T} \ , \qquad \left(\overline{D}_{\dot{\alpha}} T \ = 0 \, \right)
\eeq{32}
and ${\cal L}$ becomes
\beq
{\cal L} \ = \ \Phi(T \ + \ \overline{T} \ + \ m\, V) \ +  \ {\cal L}_{BI} \left[ g\,,\,\mu\, ,\, W_\alpha(V),  \overline{W}_{\dot{\alpha}}(V)\right] \ .
\eeq{33aaa}
This is the supersymmetric Stueckelberg representation of a massive vector multiplet. Making use of the gauge invariance of ${\cal L}_{BI}$, one can turn it into the Proca--like form
\beq
{\cal L} \ = \ \Phi(m\, V) \ +  \ {\cal L}_{BI} \left[ g\,,\,\mu\, ,\,W_\alpha(V),  \overline{W}_{\dot{\alpha}}(V)\right] \ ,
\eeq{33}
where $\Phi(mV)$ contains supersymmetric generalizations of vector and scalar mass terms, but also a scalar kinetic term. As we have stressed, the massive multiplet contains a physical scalar, which is the very reason for the presence of the arbitrary function $\Phi$. The supersymmetric Proca--like mass term that extends $A_\mu^2$ is thus generally dressed by a scalar function.

The dual supersymmetric Born--Infeld action is the supersymmetric completion of eq.~\eqref{15}. It can be obtained integrating by parts the Green--Schwarz term in eq.~\eqref{31} and then going to a first--order form for $W_\alpha$, which requires the introduction of the dual gauge field $M_\alpha(V_D)$. To begin with, however, notice that one can integrate out $U$, thus replacing the first two terms with the Legendre transform
\beq
\psi(L) \ = \ \left. \left[\Phi(U) \ - \ U \, \Phi^\prime(U)\right] \right|_{\Phi^\prime(U) = - L} \ .
\eeq{34}
Combining all these terms, the first--order Lagrangian takes the form
\beq
{\cal L} \ = \ \left. \left\{\psi(L) \ + \ {\cal L}_{BI} \left[ g\,,\,\mu\, ,\,W_\alpha,  \overline{W}_{\dot{\alpha}}\right] \right\} \right|_D  \ + \ \left. \left\{i\, W^\alpha \left[ m L_\alpha \ + \ M_\alpha(V_D) \right] \ + \ {\rm h.c.} \right\} \right|_F \ .
\eeq{35}
The notation is somewhat concise, since ${\cal L}_{BI}$ also contains an $F$--term, as we have seen in eq.~\eqref{26}.

Notice that, in this richer setting, $\psi(L)$ contains in general non--linear interactions of the massive scalar present in the massive linear multiplet, which is dual to the massive scalar of the massive vector multiplet. Both multiplets contain four bosonic degrees of freedom (a scalar and a tensor in the linear multiplet, and a scalar and a vector in the dual vector multiplet). In the massless limit the linear multiplet becomes dual to a chiral multiplet, while the vector multiplet becomes self--dual.

Integrating over $W_\alpha$ and using the self--duality of ${\cal L}_{BI}$ one finally gets
\beq
{\cal L} \ = \ \psi(L) \ + \ {\cal L}_{BI} \left[\frac{1}{g}\, , \, \frac{\mu}{g^{\,2}}\, , \, m \, L_\alpha \ + \ M_\alpha(V_D) \, ,\, m \, \overline{L}_{\dot{\alpha}} \ + \ \overline{M}_{\dot{\alpha}}(V_D) \right] \ ,
\eeq{36}
where both contributions contain a $D$--term and, as we have seen in eq.~\eqref{26}, ${\cal L}_{BI}$ also contains an $F$--term.

This implies, in particular, that the first--order non--linear correction to the linear multiplet mass term is proportional to $\left. \frac{m^{\,4}\,g^{\,6}}{\mu^{\,2}}\ L^\alpha\, L_\alpha\, \overline{L}_{\dot{\alpha}}\, \overline{L}^{\,\dot{\alpha}}\right|_D$.
Notice that for $m \neq 0$ $M_\alpha(V_D)$ can be shifted away, so that the superfield $L_\alpha$ acquires a Born--Infeld--like mass term. A three--dimensional analogue of this system was previously discussed in \cite{yer3d}.

%%%%%%%%%%%%%%%%%%%%%%%%%%%%%%%%%%%%%%%%%%%%%%%%%%%%%%%
\scs{Extended dualities for two--field Born--Infeld systems}\label{sec:9}
%%%%%%%%%%%%%%%%%%%%%%%%%%%%%%%%%%%%%%%%%%%%%%%%%%%%%%%

It is interesting to present two--field extensions of the Born--Infeld systems whose field equations enjoy an extended duality symmetry. It is well known, in fact, that the standard Born--Infeld Lagrangian possesses a $U(1)$ duality of this type, just like the Maxwell theory. This is to be contrasted with the multi--field generalizations discussed in Sections \ref{sec:4} and \ref{sec:6}, which possess a second non--linearly realized supersymmetry but typically possess no duality symmetry of this type. We would like to conclude presenting three relatively simple extensions of the Born--Infeld action to a pair of curvatures $F^1$ and $F^2$ whose field equations possess $U(1)$, $U(1)\times U(1)$ and $SU(2)$ duality symmetries. All these models solve subsets of the non--linear Gaillard--Zumino constraints \cite{gz},
\bea
&& G^i \, {\widetilde{G}}^j \ + \ F^i \, {\widetilde{F}}^j \ = \ 0 \ , \label{B4_1}\\
&& G^i \, {\widetilde{F}}^j \ - \ G^j \, {\widetilde{F}}^i \ = \ 0 \ ,
\eea{B4}
where
\beq
\widetilde{G}^i \ = \ 2 \ \frac{\partial {\cal L}}{\partial F^i} \ ,
\eeq{B6}
and in this case $i,j=1,2$.

The Lagrangian with $U(1)$ invariance reads
\beq
{\cal L} \ = \  f^2 \left[ \, 1\ -\ \sqrt{1\,+\,\frac{(F^1)^2+(F^2)^2}{f^2}
\,-\,\frac{\left(\star\left[F^1 \wedge F^2  \right]\right)^2}{f^4}}\,\right] \ ,
\eeq{B65}
and was obtained as an application of double duality between two forms of the same degree. This $U(1)$ corresponds to the constraint \eqref{B4_1} for $i=1,j=2$.

The other two examples are simply different complexifications of the two--field system, whose manifest $U(1)$ invariance corresponds to the single constraint of eq.~\eqref{B4}. In particular, letting
\beq
F \ = \ F^1 \ + \ i\, F^2 \ ,
\eeq{B66}
the case with $U(1) \times U(1)$ invariance, where the second $U(1)$ factor commutes with $SU(2)$, reads
\beq
{\cal L} \ = \ f^2 \left[ \, 1\ -\ \sqrt{1\,+\,\frac{F\cdot \overline{F}}{f^2}
\,-\,\frac{\left(\star \left[F \wedge \overline{F} \right]\right)\,\left(\star \left[{F} \wedge \overline{F} \right]\right)}{f^4}}\,\right] \ .
\eeq{B67}
This is actually a particular case of the $U(n,n)$ dualities of \cite{abmz} for $n=1$, where in the absence of scalars $U(1,1)$ reduces to its maximal compact subgroup $U(1) \times U(1)$. Finally, the case with $SU(2)$ invariance reads
\beq
{\cal L} \ = \ f^2 \left[ \, 1\ -\ \sqrt{1\,+\,\frac{F\cdot \overline{F}}{f^2}
\,-\,\frac{\left(\star \left[F \wedge F \right]\right)\,\left(\star \left[\overline{F} \wedge \overline{F} \right]\right)}{f^4}}\,\right] \ .
\eeq{B13}
where the other two constraints correspond to the $SU(2)$ generators which commute with the second $U(1)$ of the preceding example.
From these examples it is manifest that the standard Born--Infeld system admits several types of inequivalent complexifications that differ in their quartic couplings.

A fourth option, with maximal $U(2)$ duality, is naturally captured by the construction of \cite{abmz}, but its Lagrangian is not known in closed form.

\vskip 24pt

\noindent{\large \bf Acknowledgements}\\ \noindent
We  are grateful to L.~Andrianopoli, I.~Antoniadis, P.~Aschieri, A.~Ceresole, D.~Dall'Agata, R.~D'Auria, E.~Dudas, P.~Fr\'e, R.~Kallosh, N.~Kitazawa, A.~Linde, S.~Patil, M.~Porrati, A.~S.~Sorin, R.~Stora, M.~Trigiante, A.~Van Proeyen,  A.~Yeranyan and F.~Zwirner for useful discussions and/or collaborations. This work was supported in part by the ERC Advanced Investigator Grant n.~226455 (SUPERFIELDS). A.~S. is on sabbatical leave, supported in part by Scuola Normale and by INFN (I.S. Stefi), and would like to thank the CERN Th-Ph Department for the kind hospitality.
%
%%%%%%%%%%%%%%%%%%%%%%%%%%%%%%%%%%%%%%%%%%%%%%%%%%%%%%%

\end{document}